\documentclass[journal]{IEEEtran}
\usepackage[T1]{fontenc}
\usepackage[latin9]{inputenc}
\usepackage{amsmath}
\usepackage{amssymb}
\usepackage{amsthm,bm,graphicx}
\usepackage{url}
\usepackage[table]{xcolor}  
\usepackage{tabularx}
\usepackage{pifont}
\usepackage{subcaption}
\usepackage{siunitx}
\usepackage{tikz}
\usepackage{pgfplots, flushend}

\pgfplotsset{compat=newest}
\usetikzlibrary{calc, angles, quotes, scopes}

\makeatother

\usepackage{babel,verbatim,balance}
\usepackage{tcolorbox}
\usepackage{graphics, graphicx}
\usepackage{upgreek}
\usepackage{setspace,gensymb}

\begin{document}
\title{Characterization of Indoor RIS-Assisted Channels\\at $304$ \(\rm GHz \): Experimental Measurements, Challenges, and Future Directions}
\author{George C. Alexandropoulos,~\IEEEmembership{Senior~Member,~IEEE,} Bo Kum Jung, \\Panagiotis Gavriilidis,~\IEEEmembership{Graduate Student~Member,~IEEE,} S\'{e}rgio Matos, Lorenz H. W. Loeser, Varvara Elesina,\\ Antonio Clemente,~\IEEEmembership{Senior~Member,~IEEE,} Raffaele D'Errico,~\IEEEmembership{Senior~Member,~IEEE,} \\Lu\'{i}s M. Pessoa,~\IEEEmembership{Senior~Member,~IEEE,} and Thomas K\"{u}rner,~\IEEEmembership{Fellow, IEEE}
\thanks{This work has been supported by the Smart Networks and Services Joint Undertaking (SNS JU) project TERRAMETA under the European Union's Horizon Europe research and innovation programme under Grant Agreement No 101097101, including top-up funding by UK Research and Innovation
(UKRI) under the UK government's Horizon Europe funding guarantee.}
\thanks{G. C. Alexandropoulos and P. Gavriilidis are with the Department of Informatics and Telecommunications, National and Kapodistrian University of Athens, 15784 Athens, Greece. G. C. Alexandropoulos is also with the Department of Electrical and Computer Engineering, University of Illinois Chicago, Chicago, IL 60601, USA. (e-mails: \{alexandg, pangvr\}@di.uoa.gr).}
\thanks{B. K. Jung, L. H. W. Loeser, V. Elesina, and T. K\"{u}rner are with the Institute for Communications Technology, Technische Universit\"{a}t Braunschweig, 38106 Braunschweig, Germany. (e-mails: \{bo.jung, lorenz.loeser, varvara.elesina, t.kuerner\}@tu-braunschweig.de).}
\thanks{S. Matos is with Instituto de Telecomunica\c{c}\~{o}es and the Departamento de Ci\^{e}ncias e Tecnologias da Informa\c{c}\~{a}o, Instituto Universit\'{a}rio de Lisboa (Iscte-IUL), 1649-026 Lisbon, Portugal (e-mail: sergio.matos@iscte-iul.pt).}
\thanks{A. Clemente and R. D'Errico are with CEA-Leti, MINATEC Campus, 38054 Grenoble, France, and Universite Grenoble Alpes (e-mails: \{antonio.clemente, raffaele.derrico\}@cea.fr).}
\thanks{L. M. Pessoa is with INESC TEC, Campus da FEUP and the Faculty of Engineering, University of Porto, 4200-465 Porto, Portugal (e-mail: luis.m.pessoa@inesctec.pt).}
}

\maketitle
\begin{abstract}
Reconfigurable Intelligent Surfaces (RISs) are expected to play a pivotal role in future indoor ultra high data rate wireless communications as well as highly accurate three-dimensional localization and sensing, mainly due to their capability to provide flexible, cost- and power-efficient coverage extension, even under blockage conditions. However, when considering beyond millimeter wave frequencies where there exists \(\rm GHz \)-level available bandwidth, realistic models of indoor RIS-parameterized channels verified by field-trial measurements are unavailable. In this article, we first present and characterize three RIS prototypes with \( 100 \times 100 \) unit cells of half-wavelength inter-cell spacing, which were optimized to offer a specific non-specular reflection with \( 1 \)-, \( 2 \)-, and \( 3 \)-bit phase quantization at $304$~\(\rm GHz \). The designed static RISs were considered in an indoor channel measurement campaign carried out with a $304$~\(\rm GHz \) channel sounder. Channel measurements for two setups, one focusing on the transmitter-RIS-receiver path gain and the other on the angular spread of multipath components, are presented and compared with both state-of-the-art theoretical models as well as full-wave simulation results. The article is concluded with a list of challenges and research
directions for RIS design and modeling of RIS-parameterized channels at \(\rm THz \) frequencies.
\end{abstract}

\begin{IEEEkeywords}
Channel sounding, reconfigurable intelligent surfaces, THz, path loss modeling, power angular profile.
\end{IEEEkeywords}

\section{Introduction}\label{sec:intro}
The extensive unlicensed bandwidth at the \(\mathrm{THz}\) frequency band ($0.1-10$~\(\rm THz \)~\cite{ETSI_THz}) is expected to enable ultra high capacity outdoor backhaul links as well as indoor immersive and industrial applications, which will require ultra high data rates and highly accurate three-dimensional localization and sensing~\cite{VTM_THz_2024}. However, signal propagation at \(\mathrm{THz}\) frequencies suffers from high penetration losses, a fact that necessitates the design of efficient \(\mathrm{THz}\) transmitters with large output power~\cite{MML24} and/or extremely large Multiple-Input Multiple-Output (MIMO) systems optimized for very directive beamforming in the far-field and very accurate beam focusing in the usually occuring near field. The hardware complexity and power consumption of antenna arrays are though prohibitive with traditional approaches relying on phase shifters networks and multiple radio-frequency chains.

Metasurfaces constitute an emerging technology of artificial sheets comprising metallic patches or dielectric engravings of tunable responses in planar or multi-layered configurations with subwavelength thickness. Resonant effects determined by the geometry of the RIS constituent unit cells and their spatial distribution facilitate their interaction with electric and/or magnetic fields. This enables various advanced functionalities, including beam steering and shaping, flat lenses, artificial magnetic conductors, cloaking, absorbers, and scattering reduction. Furthermore, metasurfaces with unit cells of programmable responses, known as Reconfigurable Intelligent Surfaces (RISs) in wireless communications~\cite{BAL2024}, can be optimized to enhance network coverage establishing additional links bypassing signal blockages. However, designing RISs for \(\mathrm{THz}\) frequencies faces considerable fabrication and operational challenges~\cite{ACM2024}. In particular, the unit cell dimensions required for such frequencies are of the order of micrometers, demanding highly accurate fabrication processes, and the integration with switching technology remains challenging~\cite{MML24}. Additionally, configuring RISs dynamically in response to environmental changes adds complexity particularly at \(\mathrm{THz}\) where channel dynamics are rapid~\cite{VTM_THz_2024}. 

\begin{figure*}[t]
    \centering
    \includegraphics[width=0.8\textwidth]{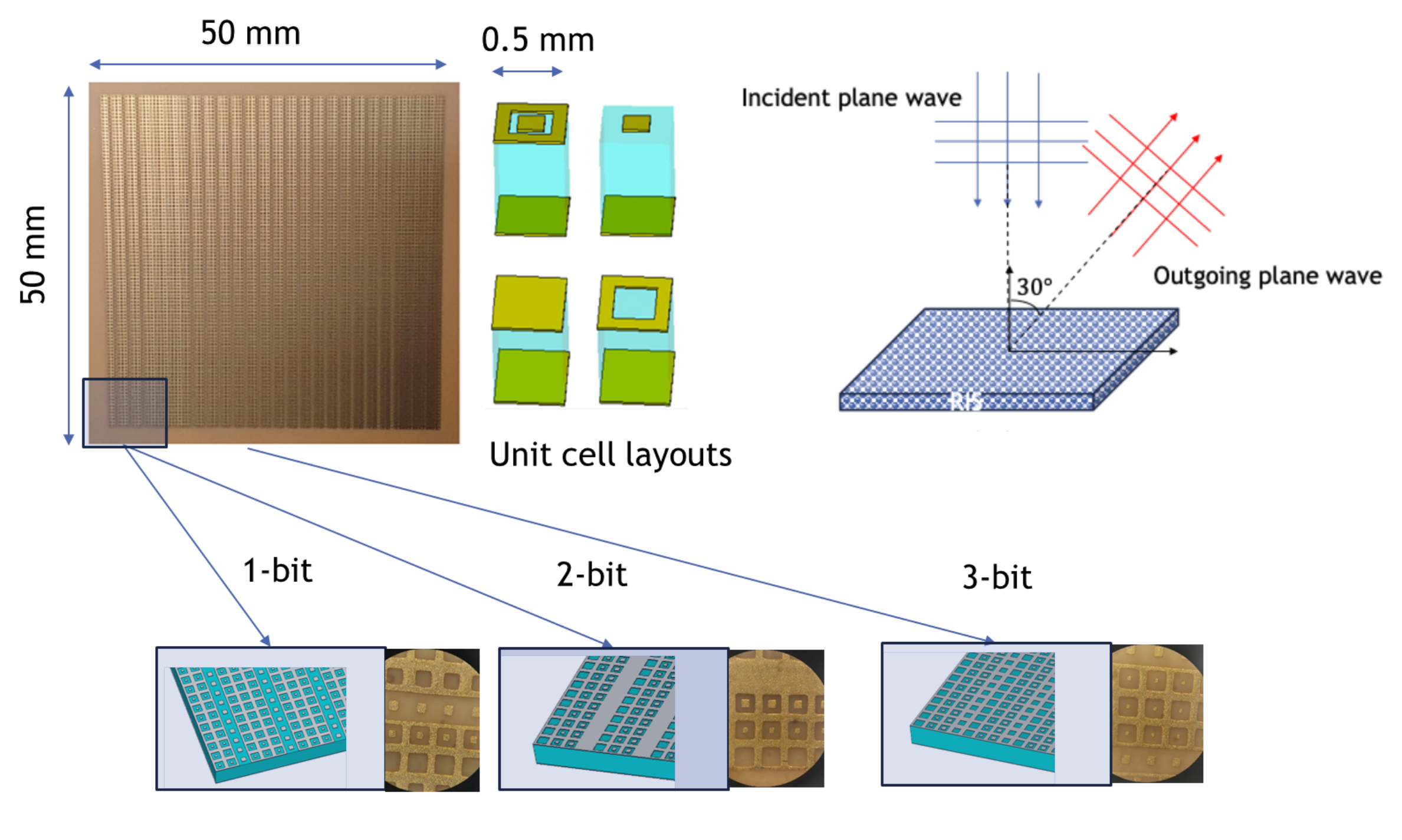}
    \caption{The fabricated static RISs at $304$~\(\rm GHz \) with \( 100 \times 100 \) unit cells of half-wavelength inter-cell spacing and \( 1 \)-, \( 2 \)-, and \( 3 \)-bit phase quantization designed to offer a non-specular reflection from a normal incidence to a \( 30^{\circ} \) outgoing plane.}
    \label{fig:R_RIS_design_123_bits}
\end{figure*}
The potential of RISs to enable \(\mathrm{THz}\) X-haul links, an alternative solution to optical fibers for connecting base stations in ultra-densified networks, were recently verified through simulations in~\cite{Jung24}. Such links require Line-of-Sight (LoS) connections, and RISs were deployed to enable virtual LoS paths, each via a single programmable non-specular reflection~\cite{BAL2024}. However, to fully understand the impact of RISs in complex \(\mathrm{THz}\) wireless channels, as appearing in dense indoor environments, channel models that accurately predict the behavior resulting from all possible RIS phase configurations, are needed. Current literature though often relies on idealistic RIS assumptions, such as unit-amplitude reflections, absence of mutual coupling, and continuous control of each cell's phase. While these assumptions were useful for initial studies, the increasing technology readiness level necessitates realistic channel models to better understand the requirements for RIS-enabled THz applications. More importantly, those models need to be grounded to channel measurements, as theoretical models alone may not fully capture the complex behavior of RISs in real-world scenarios.

In this article, we present for the first time 
the results of an indoor point-to-point channel measurement campaign at \(304\)~\(\mathrm{GHz}\) incorporating prototypes of static RISs with \( 1 \)-, \( 2 \)-, and \( 3 \)-bit phase quantization. In Section~\ref{sec:RIS_designs}, we describe and characterize the designed RISs, comparing theoretical analysis with respective full-wave Electro-Magnetic (EM) simulation results. Section~\ref{section: channel measurement campaign} introduces the two channel measurement setups and discusses the comparison between measured data and state-of-the-art theoretical models. A list of challenges and research directions for RIS design and modeling of RIS-parameterized channels at \(\rm THz \) frequencies is included in Section~\ref{sec:challenges}. Finally, the article is concluded in Section~\ref{sec:conclusions}.

\section{Design of a Static RIS at $304$ GHz}\label{sec:RIS_designs}
Three metasurfaces, each of \( 5 \times 5 \, \text{cm}^2 \) aperture including \( 100 \times 100 \) unit cells of half-wavelength inter-cell spacing, optimized to provide a fixed non-specular reflection at $304$~\(\rm GHz \) from a normal incidence to a \( 30^{\circ} \) outgoing plane wave have been designed and fabricated using low-cost Printed Circuit Board (PCB) technology, as illustrated in Fig.~\ref{fig:R_RIS_design_123_bits}. The desired reflection phase profile has been first optimized offline considering elements each with \( 1 \)-, \( 2 \)-, and \( 3 \)-bit phase quantization capability, and then, the respective static unit cells of the three RISs were designed and fabricated. The layouts of the unit cells as well as the designs and prototypes for all three phase quantization cases are also shown in the figure.
Note that, besides the prescribed \( 30^{\circ} \) outgoing wave (main beam), the scattered field from the RIS will produce two additional outgoing wave components: the specular reflection in the normal direction and a mirror beam at \( -30^{\circ} \). The amplitude ratio between the main beam and the mirror beam depends on the considered phase quantization resolution. Indicatively, Table~\ref{tbl:full_wave_RCS_123_bits_R_RIS} shows that, for \( 1 \)-bit resolution, the two beams share the same Radar Cross Section (RCS) value, while, for \( 2 \)-bit, the ratio is \( 12\)~\(\rm dB \), and, for \( 3 \)-bit, it is \( 15\)~\(\rm dB \). This table includes also the RCS values for the specular reflection, as well as the aperture efficiency defined as the ratio between the RCS at the main beam and that of a Perfect Electric Conductor (PEC) plane of the same size under normal incidence and specular reflection. The larger the aperture efficiency is, the more efficient is the designed static RIS.

\begin{table*}[t]
\centering
\caption{RCS values and aperture efficiencies of the designed RISs with \( 1 \)-, \( 2 \)-, and \( 3 \)-bit phase quantization.}\label{tbl:full_wave_RCS_123_bits_R_RIS}
\resizebox{2\columnwidth}{!}{%
\begin{tabular}{|c|c|c|c|c|}
\hline
\textbf{Quantization Bits} & \textbf{RCS} at \(30^{\circ}\) ($\rm dBsm$)& \textbf{RCS} at \(-30^{\circ}\) ($\rm dBsm$)& \textbf{RCS} at \(0^{\circ}\) ($\rm dBsm$)& \textbf{Aperture Efficiency} \\
\hline\hline
\(3\)  & 16.7 & 1.7 & -2.6 & 59\% \\
\hline
\(2\) & 15.7 & 3.7 & 6.1  & 47\% \\
\hline
\(1\)  & 12.2 & 12.2 & 1.1  & 21\% \\
\hline
\end{tabular}%
}
\end{table*}

\begin{figure}[t]
    \centering
    \includegraphics[width=\columnwidth]{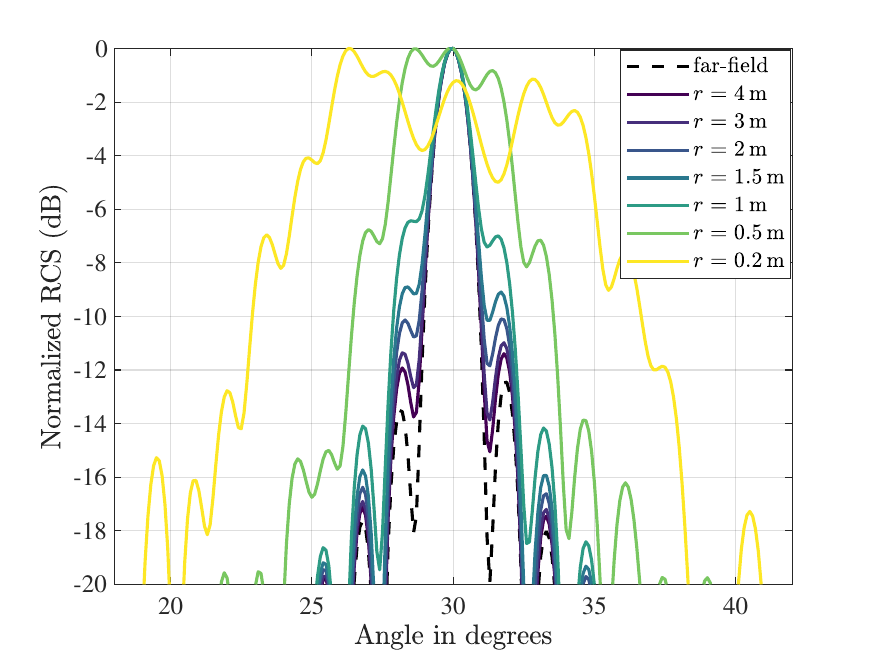}
    \caption{Normalized radiation pattern of the designed \( 100 \times 100 \) static RIS with \(3\)-bit phase resolution with respect to the RCS of the main beam for different distances \(r\) from the RIS center.}
    \label{fig:R_RIS_near_field_patterns_decreasing_distances}
\end{figure}

The radiation pattern of the designed \(3\)-bit static RIS with  \( 100 \times 100 \) unit cells, as obtained through full-wave EM simulations for different distances \(r\) from the center of the metasurface, is illustrated in Fig.~\ref{fig:R_RIS_near_field_patterns_decreasing_distances}. Emphasis was given in \(r\) values within the near-field region of this RIS, which, following the definitions of the Fresnel distance and the Rayleigh Distance (RD)~\cite{near_field_review}, spans the range from \(0.2\, \mathrm{m}\) to \(r_{\mathrm{RD}}=10.15\, \mathrm{m}\) with respect to the RIS center. It can be observed that, even without explicitly considering near-field effects during the RIS design, there exist \(r\) values below $r_{\mathrm{RD}}$, where efficient radiation patterns occur. By setting the criteria for satisfactory beam patterns in the near-field region as: \textit{i}) the Half-Power Beam Width (HPBW) remains unchanged, and \textit{ii}) the normalized Side Lobe Level (SLL) is below \(-10\, \mathrm{dB}\), it can be inferred from the figure that the RIS operates efficiently at distances greater than \(1.5\, \mathrm{m}\), which is significantly below the Rayleigh distance $r_{\mathrm{RD}}$. In fact, we have confirmed that the simulated radiation patterns are almost identical for \(r>4\, \mathrm{m}\).

Interestingly, the findings from Fig.~\ref{fig:R_RIS_near_field_patterns_decreasing_distances} are consistent with the existing literature on beam focusing performance for planar arrays~\cite{near_field_Bjornson,near_field_beam_tracking}. Specifically, the boundary between far- and near-field communications is delineated by the distance \(r_{\mathrm{DF}}\) related to the Depth of Focus (DF), which marks the transition where beam focusing, rather than beam steering, should be considered for beam design. For \(r > r_{\mathrm{DF}}\), beam steering causes less than \(3\, \mathrm{dB}\) gain error, whereas, for \(r < r_{\mathrm{DF}}\), gain differences exceed that value. For the designed fixed RIS at $304$~\(\rm GHz \) with \( 100 \times 100 \) unit cells of half-wavelength inter-cell spacing, it holds that \(r_{\mathrm{DF}} \approx 0.1\, r_{\mathrm{RD}} = 1.15\, \mathrm{m}\)~\cite{near_field_Bjornson}. This value closely matches the limiting distance of \(1.5\, \mathrm{m}\) inferred from Fig.~\ref{fig:R_RIS_near_field_patterns_decreasing_distances} using the HPBW and SLL criteria. It is noted, however, that the \(1.15\, \mathrm{m}\) distance has been derived for the specific case where the receiver lies along the direction dictated by the normal vector originating from the RIS center, hence, it does not apply for other cases. To this end, to compute the distance from which the error of the far-field beam steering will be larger than \(3\) dB, whilst taking into account the \(30^{\circ}\) tilt of the beam, the more general formula proposed in \cite[Lemma~1]{near_field_beam_tracking} can be employed which results in the limiting distance \(r_{\mathrm{DF},30^{\circ}} = 0.7 \,\mathrm{m}\). 

Figure~\ref{fig:R_RIS_near_field_patterns_decreasing_distances} depicts that the RIS radiation patterns are almost identical for distances \(r>1\, \mathrm{m}\), confirming a rough agreement between full-wave EM simulations and theoretical analysis~\cite{near_field_Bjornson,near_field_beam_tracking}. While performance analysis predicts a \(3\, \mathrm{dB}\) difference in the main beam's gain between the far-field and the case of \(r = 0.7\, \mathrm{m}\), such differences are not observed in the simulations for \(r>0.5\, \mathrm{m}\). Discrepancies between theoretical predictions and full-wave simulations may arise from several factors, including the assumption of continuous phase resolution and ideal unit amplitude reflections in theoretical models versus the \(3\)-bit phase quantization and the \(59\%\) aperture efficiency of the designed static RIS. To refine theoretical predictions, effective aperture can replace physical aperture in calculating the distance where \(3\, \mathrm{dB}\) beamforming error occurs. To this end, using the formula in \cite[Lemma~1]{near_field_beam_tracking}, the error distance is \(r = 0.41\, \mathrm{m}\). This refinement aligns better with simulations, where an \(1.5\, \mathrm{dB}\) error is shown at \(r = 0.2\, \mathrm{m}\), versus the theoretical \(3\, \mathrm{dB}\) at \(0.41\, \mathrm{m}\). Motivated by these findings, in the following Section~\ref{section: channel measurement campaign}, we compare the theoretical beamforming error predictions using the effective aperture against experimental measurements.

\section{Indoor Channel Measurements at $304$ GHz}\label{section: channel measurement campaign}
In this section, we present channel sounding results at \(304 \, \mathrm{GHz}\) in a typical indoor setting incorporating the designed RISs with the goal to enhance the received signal strength.

\begin{figure*}[tb!]
    \centering
    \begin{subfigure}[b]{0.48\textwidth}
        \centering
        \includegraphics[width=\textwidth]{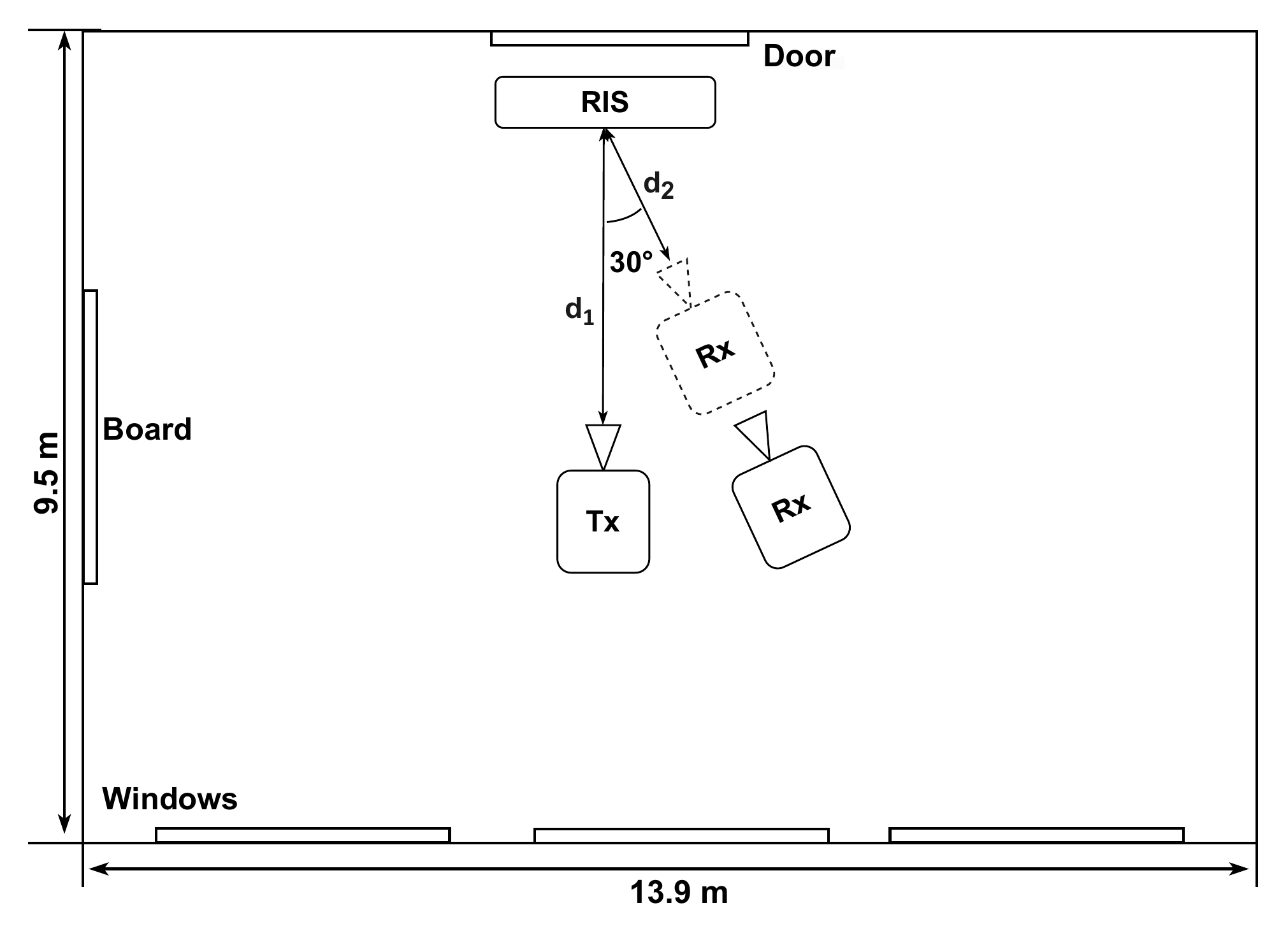} 
        \caption{Setup 1: Path gain measurements.}
        \label{fig: Schematic view of P2P measurement}
    \end{subfigure}
    \hfill
    \begin{subfigure}[b]{0.48\textwidth}
        \centering
        \includegraphics[width=\textwidth]{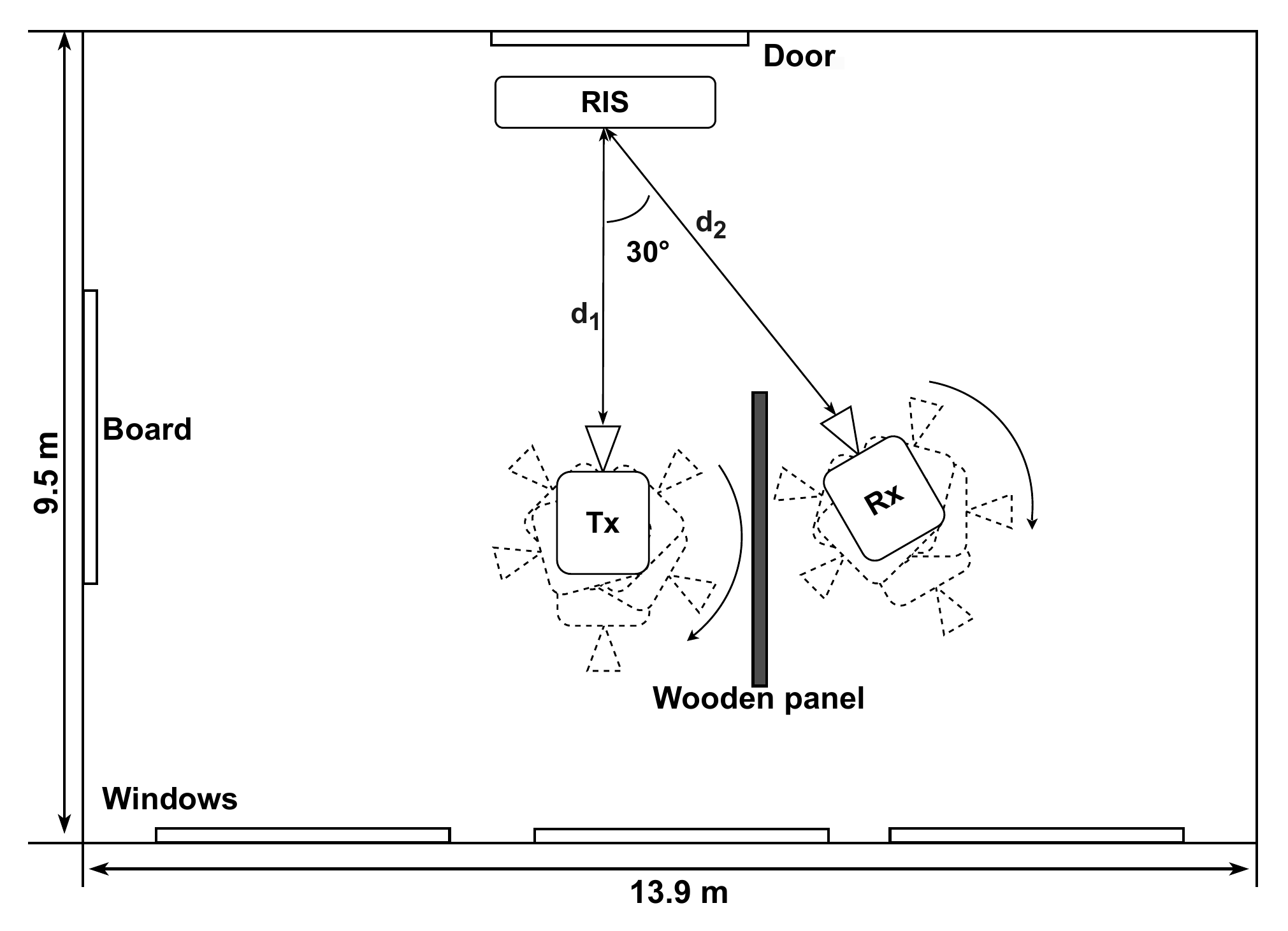} 
        \caption{Setup 2: Power angular profile measurements.}
        \label{fig: Schematic view of angular sweep measurement}
    \end{subfigure}
    \caption{Schematic view of the indoor channel measurements setups at $304$ GHz incorporating the designed static RISs.}
    \label{fig:largefigure}
\end{figure*}

\subsection{Measurement Setups}
A channel sounder at \(304 \, \mathrm{GHz}\) from Technische Universit{\"a}t Braunschweig~\cite{TUBS_ChannelSounder_Speficification} was employed to measure the Channel Impulse Response (CIR) in the time domain using correlated ultra-wide band pseudo-random signals. Detailed specifications of the channel sounder are provided in Table~\ref{Tab: channel sounder parameters}. The measurements took place in a lecture room with dimensions \(13.9~\mathrm{m}~\times~9.5~\mathrm{m}\) including one Transmitter (Tx) and one Receiver (Rx) with horn antennas, and either of the designed static RISs, as shown in Fig.~\ref{fig:largefigure}. Two different setups were realized to characterize the indoor RIS-assisted wireless channel: one focusing on end-to-end path gain measurements and the other on measurements for the channel's Power Angular Profile (PAP).

The measured data of the gain of the end-to-end Tx-RIS-Rx channel path from the first setup depicted in Fig.~\ref{fig: Schematic view of P2P measurement} were compared with both a conventional and a state-of-the-art path gain formula. In this setup, the Tx, Rx, and RIS were all placed on tripods at height \(1.52\,\mathrm{m}\), whereas the two former were centrally positioned in the room, while the RIS was located close to one of the surrounding walls. The Tx faced the RIS perpendicularly and the Rx was positioned at the angle \(30^{\circ}\) relative to RIS's normal vector. The Tx-RIS distance was fixed at \(d_1=2.15\,\mathrm{m}\), while the Rx-RIS distance varied from \(d_2=0.2~\mathrm{m}\) to \(10~\mathrm{m}\). The setup aimed to investigate the RIS beamforming capability to enable Tx-Rx communications, when their directive antennas were only aligned through the RIS (i.e., highly attenuated direct Tx-Rx LoS channel).

The second setup illustrated in Fig.~\ref{fig: Schematic view of angular sweep measurement} was similar to the first one, except that also the Rx was fixed at the distance \(d_2=5.5\,\mathrm{m}\) and the direct LoS path between Tx and Rx was obstructed by a wooden panel, thus, emulating a LoS blockage scenario. The goal of this setup was to measure the angular spread and strength of the MultiPath Components (MPCs) of the channel~\cite{EuCAP2025_angular_sweep_measurement}. This investigation enabled the identification of their directional characteristics, such as the Angles of Arrival (AoAs) and Angles of Departure (AoDs). This information is especially valuable for understanding the MPC propagation characteristics of a given environment, thus, enabling more robust MIMO beamforming designs. 
\begin{table}[tb!]
\centering
\caption{Parameters of the  \(304 \, \mathrm{GHz}\) channel sounder.}
\begin{tabular}{|l|l|}
\hline
\textbf{Parameter}        & \textbf{Value}        \\ \hline\hline
Center frequency & 304.2 GHz    \\ \hline
Clock frequency  & 9.22 GHz     \\ \hline
Bandwidth        & 8 GHz        \\ \hline
Chip duration    & 108.5 ps     \\ \hline
Sequence length  & 4095         \\ \hline
Sequence duration  & 444.14 ns   \\ \hline
Sampling factor  & 128          \\ \hline
Measurement rate & 17,590 CIR/s \\ \hline
Tx and Rx antenna gain & 26.4 dBi \\ \hline
Tx and Rx antenna HPBW  & 8.5$^{\circ}$ \\ \hline
\end{tabular}
\label{Tab: channel sounder parameters}
\end{table}

\subsection{Setup 1: Path Gain Measurements}
The RCS values, $\sigma_{\mathrm{RIS}}$, at \( 30^{\circ} \) for the designed static RISs included in Table~\ref{tbl:full_wave_RCS_123_bits_R_RIS} can be substituted into the following bi-static far-field radar formula to calculate the free-space path gain of the end-to-end Tx-RIS-Rx path for the setup in Fig.~\ref{fig: Schematic view of P2P measurement}  ($\lambda$ is the wavelength, in our case, for the frequency \(304 \, \mathrm{GHz}\)):
\begin{equation}
\label{bistatic radar equation}
P_{\rm ff} = \frac{\sigma_{\mathrm{RIS}}\lambda^2}{(4\pi)^3 d_1^2 d_2^2}.
\end{equation}
It is noted, however, that all RISs in Section~\ref{sec:RIS_designs} were designed using the far-field approximation. They were particularly optimized to reflect the Tx signal at \( 30^{\circ} \) with respect to the RIS center in RIS-Rx distances within the far-field region. This fact introduces a mismatch with the actual Rx position with polar coordinates \((d_2, 30^{\circ})\), which will impact RIS's beamforming gain in the near-field. To account for this mismatch in the theoretical computation of the path gain, we adopt the analytical model proposed very recently in~\cite{near_field_beam_tracking}, which quantifies such beamforming errors. To this end, the Tx-RIS-Rx path gain in the near-/far-field is given by: 
\begin{equation}
\label{bistatic radar equation adjusted}
P_{\rm nf,ff} =  \mathcal{K}^2\left(d_2,30^{\circ}\right)P_{\rm ff},
\end{equation}
where, for the computation of function $\mathcal{K}\left(\cdot,\cdot\right)$ \cite[Lemma~1]{near_field_beam_tracking} that captures the beamforming error, we have used the RIS effective aperture reported in Table~\ref{tbl:full_wave_RCS_123_bits_R_RIS} for all three designs.

\begin{figure}
    \centering
    \includegraphics[width=\linewidth]{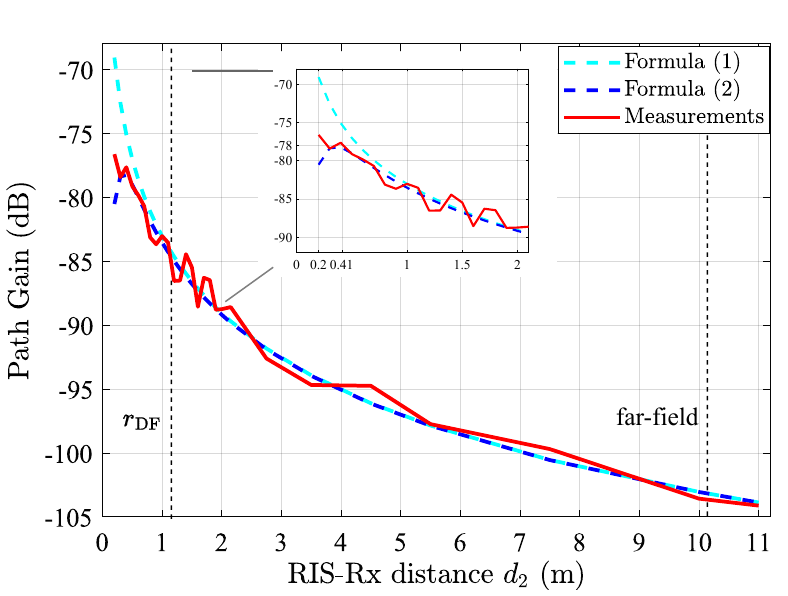}
    \caption{Comparison of the measured and theoretical (both formulas~\eqref{bistatic radar equation} and~\eqref{bistatic radar equation adjusted}) gains of the end-to-end Tx-RIS-Rx path with the designed $3$-bit RIS for the Setup 1 in Fig.~\ref{fig: Schematic view of P2P measurement} versus the RIS-Rx distance $d_2$.}
    \label{fig: Point to Point Measurement Path Gain at 30deg}
\end{figure}

Figure~\ref{fig: Point to Point Measurement Path Gain at 30deg} illustrates the gain of the end-to-end Tx-Rx path via the designed $3$-bit RIS for Setup 1 as a function of the RIS-Rx distance $d_2$. To account for the systematic measurement errors across all considered $d_2$ values, which arised due to a combination of factors (e.g., channel sounder instability, misalignment, fabrication imperfections, beam squint, and unwanted reflections from setup components such as the RIS holder), a constant correction of \(5.5\,\mathrm{dB}\) has been applied to the measured results~\cite{LMK2025}. It can be observed that, for \(d_2 \geq r_\text{DF} \approx \SI{1.15}{\meter}\) (almost one tenth of the Rayleigh distance between the RIS and Rx), the measured path gain closely matches the theoretical values obtained via both the far-field and the near-/far-field formulas \eqref{bistatic radar equation} and \eqref{bistatic radar equation adjusted}, respectively. However, for \(d_2 < \SI{1.15}{\meter}\), there exists a gap between measurements and the theoretical values predicted by~\eqref{bistatic radar equation} which increases with decreasing $d_2$. This particularly indicates that~\eqref{bistatic radar equation} overestimates the path gain and this becomes more pronounced the smaller $d_2$ becomes. Interestingly, the latter does not appear with the predicted values from the near-/far-field formula \eqref{bistatic radar equation adjusted}, where excellent agreement with measured data is exhibited for \(d_2 \geq \SI{0.25}{\meter}\), thus, validating the near-field analysis in~\cite{near_field_beam_tracking}.

 \begin{figure*}[!t]
    \centering
    \begin{minipage}{\textwidth}
        \centering
        \text{Path Gain (dB)}\\
        \includegraphics[width=0.6\textwidth]{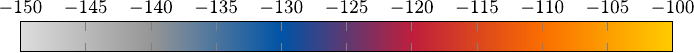}
    \end{minipage}

    \vspace{1em} 

  \begin{minipage}{\textwidth}
        \centering
        \begin{subfigure}{0.3\textwidth}
            \centering
            \includegraphics[width=\textwidth]{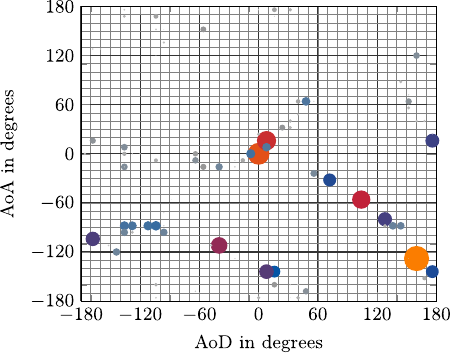}
            \caption{\(1\)-bit RIS.}
            \label{fig: angular sweep 1 bit}
        \end{subfigure}
        \hfill
        \begin{subfigure}{0.3\textwidth}
            \centering
            \includegraphics[width=\textwidth]{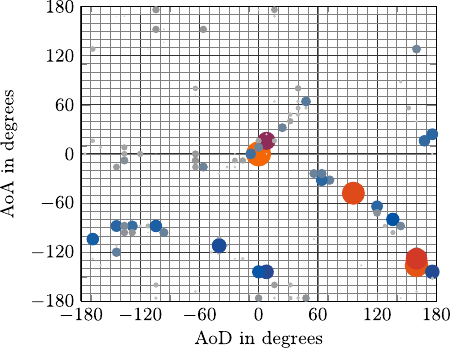}
            \caption{\(3\)-bit RIS.}
            \label{fig: angular sweep 3 bit}
        \end{subfigure}
        \hfill
        \begin{subfigure}{0.3\textwidth}
            \centering
            \includegraphics[width=\textwidth]{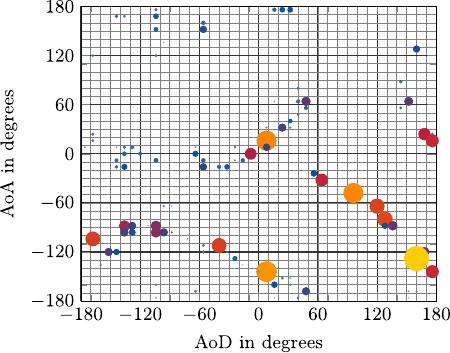}
            \caption{Backflip of an RIS (PEC).}
            \label{fig: angular sweep pec}
        \end{subfigure}
    \end{minipage}

    \caption{Power angular profile measurements for the Setup 2 illustrated in Fig.~\ref{fig: Schematic view of angular sweep measurement} considering three different cases for the RIS.}
    \label{fig: angular sweep measurement}
\end{figure*}
The path gain measurements in Fig.~\ref{fig: Point to Point Measurement Path Gain at 30deg} align well with the theoretical analysis and simulation results presented in Section~\ref{sec:RIS_designs}. Therein, full-wave EM simulations predicted a degradation in the RIS beamforming performance for distances \(d_2<\SI{1.5}{\meter}\), primarily due to increased SLL and broader HPBW. Note that, while the latter metrics-- derived from radiation pattern characteristics --cannot be directly mapped to the experimental path gain results, the observed path gain measurements differ from the far-field formula~\eqref{bistatic radar equation} at shorter distances (i.e., \(d_2 < \SI{1.5}{\meter}\)), suggesting an indirect connection. For instance, elevated SLL may contribute to the reduced path gain by redistributing power across unintended directions. On the other hand, the RIS theoretical analysis predicted a \(3 \, \mathrm{dB}\) gain discrepancy at \(d_2=\SI{0.41}{\meter}\) when considering the effective aperture of the metasurface. This aligns with the experimental observations in Fig.~\ref{fig: Point to Point Measurement Path Gain at 30deg}, where a similar \(2.8 \, \mathrm{dB}\) deviation is evident between the measured path gain and the conventional far-field formula~\eqref{bistatic radar equation}, which does not account for near-field beamforming errors.

\subsection{Setup 2: Power Angular Profile Measurements}
Three sets of measurements were conducted for this setup illustrated in Fig.~\ref{fig: Schematic view of angular sweep measurement}, each considering one of the following designed RIS prototypes: 1) $1$-bit RIS, 2) $3$-bit RIS, and 3) the backflip of any of the RISs which is a PEC, serving as an extreme case of a specular reflector. The initial directions of the Tx and Rx were set to directly face the RIS, thus, at 0$^{\circ}$ and 30$^{\circ}$ relative to the RIS, respectively. 

The PAP values of all measurement sets, which demonstrate the spatial distribution of the path gain of the different MPCs in the angular domain, are included in Fig.~\ref{fig: angular sweep measurement}. The coordinate system was defined with respect to the AoA and AoD of the initial measurement direction. Positive values for AoAs/AoDs correspond to clockwise rotation, while the negative ones to counter-clockwise rotation. An AoA/AoD value of $0^{\circ}$ indicates that the respective MPC is included in the path going directly from the Tx to Rx via the RIS (i.e., when both Tx and Rx are perfectly aligned with the RIS). It is noted that the MPCs at/from the same AoA/AoD combine their amplitudes by superposition taking into account the phase information of each individual MPC. To this end, the size of each individual circle in all subfigures represent the relative strength of the MPCs, with this strength ranging between the strongest and the weakest MPC within the specific measurement set. The absolute strength of the MPCs in dB is indicated by their color (color bar at the top of the figure).

For all measurement sets, a total of four dominant MPCs were measured: 1) the reflection from the RIS at AoD and AoA of $0^{\circ}$, 2) the obstructed direct path between Tx and Rx with AoD $95^{\circ}$ and AoA $-57^{\circ}$, 3) the reflection from the wall behind the RIS with AoD $13^{\circ}$ and AoA $17^{\circ}$, and 4) the reflection of the opposite wall of the RIS with AoD $160^{\circ}$ and AoA $-130^{\circ}$. This fact indicates that the inclusion of a static RIS or a PEC in the considered lecture room at Technische Universit{\"a}t Braunschweig results in a highly scattering environment. Interestingly, it is demonstrated in Fig.~\ref{fig: angular sweep measurement} that the RIS reduces the strength of the scattered signals as compared to the use of a PEC, and this happens due to former's radiation pattern optimization with respect to the Tx and Rx placement. As expected, when the $3$-bit RIS was deployed, the strongest MPC was the one reflected from it. Surprisingly, this was not the case when the $1$-bit RIS was used. In that case, the strongest MPC was the one reflected from the opposite wall of the RIS ( at the bottom of the figure).

By comparing the PAPs between the $1$- and $3$-bit RIS cases, it can be concluded that the deployment of the higher phase resolution metasurface results in larger spatial diversity. This implies that an RIS with more refined beamforming properties contributes more distinguishable MPCs (i.e., a larger RCS compensates more efficiently the high path loss), and thus to larger angular spread. It should be noted, however, that a direct comparison between the main and spurious beams is challenging with results obtained from a channel sounding process, since their time dependency is not captured. 

\section{Open Challenges and Future Directions}\label{sec:challenges} 
In this section, we list challenges and resulting research directions for RIS design and modeling of RIS-parameterized channels at \(\mathrm{THz}\) frequencies.

\subsection{RIS Fabrication at THz}
The dimensions of RIS unit cells at \(\mathrm{THz}\) are of the order of micrometers, making it challenging to integrate excitation circuitry at such a fine scale~\cite{MML24,ACM2024}. In addition, even minor imprecisions in the fabrication of unit cells can drastically affect their performance. For instance, for the RIS design presented in Section~\ref{sec:RIS_designs}, our full-wave EM simulations showcased that a deviation of just \(\SI{5}{\micro\meter}\) results in a phase error of up to \(25^\circ\). Given that \(\SI{20}{\micro\meter}\) is a typical precision limit for conventional PCB manufacturing, such discrepancies cannot be overlooked. Addressing these fabrication challenges is crucial for ensuring accurate RIS performance at \(\mathrm{THz}\).

\subsection{Channel Measurements with Multi-Functional RISs}
Efficient signal propagation at \(\mathrm{THz}\) requires extremely directive beamforming~\cite{near_field_beam_tracking}. Transmissive RISs constitute a recent metasurface technology for transmitter radio-frequency front-ends capable of offering exceptional directivity. In~\cite{TRIS37dBi}, such an RIS-based transmitter operating at $300$~\(\rm GHz \) was designed achieving an HPBW of approximately \(2^{\circ}\), a peak aperture close to $50.0\%$, and a wideband behavior (specifically, a relative bandwidth of $19.0\%$). It is thus interesting to conduct \(\mathrm{THz}\) channel measurement campaigns incorporating a transmissive RIS and one or more RISs with programmable reflective phase profiles, in contrast to the designed RISs with fixed profiles investigated in Section~\ref{section: channel measurement campaign}, in both indoor and outdoor environments. Campaigns in complex real-world settings with rough surfaces and complex geometries will contribute in the characterization of diffuse scattering from surface irregularities, as well as the RIS interactions with various building materials. In addition, the relationships among specific RIS phase configurations and corresponding channel parameters will be better understood. Last but not least, performing measurement with wideband \(\mathrm{THz}\) transmissions will help characterizing the beam squint effect that can distort beam patterns across frequencies~\cite{HYZ23}, thus, potentially degrading communication performance. 

As observed from Fig.~\ref{fig: angular sweep measurement}, in the backflip scenario which is equivalent to a PEC surface, more channel paths are present. This suggests a potential increase in the spatial degrees of freedom of the channel, a characteristic typically advantageous for wireless communications. However, this fact does not diminish the effectiveness of the RIS, but highlights the high directivity of our fixed reflection design. For alternative beam pattern designs, such as those optimized for sum-rate maximization in multi-user scenarios, the multipath profile of the RIS-parameterized environment could be further enhanced. Even in the current case where the RIS maximizes the gain towards a specific direction, its advantage over a PEC surface lies upon its ability to dynamically reconfigure the dominant spatial path, which is particularly useful for applications like user tracking. Interestingly, for the designed \(3\)-bit RIS prototype, the strongest MPC observed in Fig.~\ref{fig: angular sweep 3 bit} corresponds to its main reflection beam.

\subsection{Efficient Models for RIS-Parameterized Channels}
Although ray tracing is a widely accepted method for channel modeling/characterization, it is usually computationally prohibitive requiring re-execution for every minor environmental change, including any new RIS phase profile. Efficient algorithms that leverage prior ray tracing results for reducing computational complexity are thus of paramount importance. Another way to move forward could be to utilize simplified, yet accurate, models capturing the behavior of an RIS using few ray approximations, refined by its radiation pattern, to capture key propagation characteristics with minimal computational overhead.

Conducting full-wave EM simulations for extracting the RIS radiation pattern for every possible incident angle and configuration of its unit cells is computationally infeasible, especially for large RISs with multiple phase states per cell. This highlights the need for accurate, physics-compliant models for the tunable responses of the RIS unit cells. Parametric models that focus on the circuit representation of such elements, effectively capturing their EM properties, could be adapted to reduce computational overhead. Such models would not only enhance the accuracy of channel characterization, but will also support the development of practical algorithms for RIS configuration.

\subsection{Characterizing RIS-Enabled Wave Propagation Control}
The reconfigurable behavior of RISs is time-dependent, hence, developing effective time-domain channel models to capture their impact on the delay and Doppler spreads is critical. Such models can enable novel channel control strategies, including optimizing RISs to either extend channel duration by enhancing delayed paths, or reduce it by focusing on dominant channel components. Those models must also account for the RIS actuation time, since the coherence time of \(\mathrm{THz}\) channels can be on the order of microseconds~\cite{VTM_THz_2024}, thus, comparable to the RIS phase profile switching speeds~\cite{MML24}. To ensure dynamic adaptability, the RIS reconfiguration time must be optimized, otherwise, this risks exceeding the channel coherence time, thereby limiting RIS's effectiveness.



RISs are also capable to induce frequency selectivity in wireless channels. In particular, when the RIS switching speed exceeds the channel frame duration, rapid changes in the RIS radiation profile introduce intentional frequency selectivity. On the other hand, while the RIS dynamically adjusts its radiation pattern, the channel may experience varying reflection states, leading to unintended frequency selectivity. This transient behavior can either be harnessed via advanced frequency selective beamforming or require mitigation to prevent adverse effects. An initial study on the end-to-end time-domain channel behavior with RISs  presented in~\cite{ICC_2025_time_domain_RIS} remains yet to be experimentally validated.

\section{Conclusions}\label{sec:conclusions}
In this article, we first introduced static RIS designs at $304$~\(\rm GHz \) with \( 1 \)-, \( 2 \)-, and \( 3 \)-bit phase quantization, which were considered in an indoor channel measurement campaign carried out with a $304$~\(\rm GHz \) channel sounder. Measurements for two setups, one focusing on characterizing the Tx-RIS-Rx path gain and the other on the angular spread of the channel's MPC, were presented and compared with theoretical and full-wave EM simulation results. It was demonstrated that the path gain predicted from state-of-the-art analytical formulas matches very well with measured data. It was also shown that, although RISs contribute additional scattering in indoor settings, their phase configuration optimization can reduce the amount of signal leakage at unwanted directions.
 
\bibliographystyle{IEEEtran}
\bibliography{references}

\end{document}